\documentclass[12pt,a4]{article}
\begin{document}
\setlength{\textheight}{7.4in}
\setlength{\topmargin}{-0.6in}
\setlength{\oddsidemargin}{0.2in}
\setlength{\evensidemargin}{0.5in}
\renewcommand{\thefootnote}{\fnsymbol{footnote}}

\begin{center}
{\Large {\bf Once again: Instanton method vs. WKB}}

\vspace{0.8cm}

{H.J.W. M\"uller--Kirsten,\raisebox{0.8ex}{\small a}\footnote[1]
{ Email:mueller1@physik.uni-kl.de}
Jian--zu Zhang,\raisebox{0.8ex}{\small a,b}\footnote[2]
{Email:jzzhang@physik.uni-kl.de, jzzhangw@online.sh.cn}
and Yunbo Zhang\raisebox{0.8 ex}{\small a,c}\footnote[3]
{Email:ybzhang@physik.uni-kl.de}}

\raisebox{0.8 ex}{\small a)}{\it Department of Physics, University
of Kaiserslautern, 67653 Kaiserslautern, Germany}

\raisebox{0.8 ex}{\small b)}{\it  Institute for Theoretical Physics,
Box 316, East China University of Science
and Technology, Shanghai 200237, P.R. China}

\raisebox{0.8 ex}{\small c)}{\it Department of Physics and
Institute for Theoretical Physics, Shanxi University, Taiyuan, Shanxi 030006,
 P.R. China}

\end{center}

\vspace{0.3cm}

{\centerline{\bf Abstract}}
\noindent
A recent analytic test of the instanton method
performed by comparing the exact spectrum of the
Lam${\acute e}$ potential (derived from representations of a
finite dimensional matrix expressed in terms of $su(2)$
generators) with the results of the tight--binding and  instanton
approximations as well as the standard WKB approximation
is commented upon. It is pointed out  that in the case
of the Lam${\acute e}$ potential as well as
others the WKB--related method of matched
asymptotic expansions yields the exact instanton
result as a result of boundary
conditions imposed on wave functions which are matched in 
domains of overlap.

\vspace{2cm}

\section{Introduction}
We comment on ref.\cite{1}.
Ever since the instanton method of evaluating path 
integrals became popular, the question was raised
as to whether this method and that of the Schr\"odinger equation
lead to identical results in the dominant
approximation.  In view of the significance of
both methods this is an important question. Generally
one would argue
that in the 1--loop approximation of the
path integral the results of both methods
should agree -- and, in fact,
this is the widely accepted opinion.  However, there are
few models which allow explicit verification, and some detailed
investigations of even
these are not well known.  Thus recently in a study of Lam${\acute e}$
 instantons \cite{1}, a parallel consideration of the
standard WKB method (i.e. that with linear connection
relations) was found to lead to a  result which is
off by an overall factor of $\sqrt{e/\pi}$  --
very similar to an observation made long ago in ref.\cite{2} in the case of
the well known double well potential.  In the following we point
out that in both cases, the WKB method  employed is too
simple, and complete agreement between
the result of the path integral method and that of
the Schr\"odinger equation
in all the examples referred to can be achieved with the
method of matched asymptotic expansions
 \footnote{The source of the discrepancy can be traced to a
mismatch between the normalization of the WKB
wave functions and the harmonic oscillator wave functions, as
explained in ref. \cite{3}, 
and also commented on
in ref. \cite{4}.
We thank the referee for pointing this out.
Our solutions involving the function $B[z(u)]$
in terms of Hermite functions
correspond to the harmonic oscillator wave functions
of ref. \cite{3}. Matching WKB- or WKB-like functions
(like our functions involving $A(u)$)
to these therefore avoids the mismatch}.

The best known examples
for such considerations are the scalar theory with the double well
potential and the
sine--Gordon theory of the periodic potential.  The case considered
in ref.\cite{1} with the Lam${\acute e}$ potential
 is not so well known but has recently found
widespread consideration in various contexts, such as
supersymmetric quantum mechanics \cite{5}
and spin tunneling (cf. e.g. \cite{6}), and includes in a particular limit
the sine-Gordon case.  In the following
 we sketch the main points of the method and 
recall the level splitting calculated long ago for the
Lam${\acute e}$ (and more generally ellipsoidal) wave
equation \cite{7} even for an arbitrary
excited state,  and consider limits and special cases to
demonstrate the agreement with the result of the
instanton method, or more generally the
path integral method,  in particular that 
of the ground state case of ref.\cite{1}.
For details of the calculations we refer to the
literature, where many details are given.

\section{The Lam${\acute e}$ level splitting and consequences}
The Lam${\acute e}$ equation \cite{8}
\begin{equation}
y^{\prime\prime}+\bigg\{\Lambda -\kappa^2 sn^2u\bigg\}y=0, \;\; \kappa^2
=n(n+1)k^2,
\label{1}
\end{equation}
with elliptic modulus $|k|<1$ and $n>-1/2, 0<u<2{\cal K}$, can be
looked at as a Schr\"odinger equation with periodic potential
$\kappa^2sn^2u$, where $sn u$ is one of the Jacobian elliptic functions
with period $2{\cal K}$, and ${\cal K}$ is one of the complete
elliptic integrals. (In the comparison with the Schr\"odinger
equation, the usual factor $-\hbar^2/2m$ in front of the second
derivative has to be kept in mind, where $m$ is the mass).

We recapitulate briefly the main steps in the method
of matched asymptotic expansions (of wave functions)
as applied to the periodic Lam${\acute e}$ equation (\ref{1})
(for the simplest application of the method see ref. \cite{9}).
The first step is to write the eigenvalue $\Lambda $
as \begin{equation}
\Lambda (q,\kappa):=q\kappa +\frac{\triangle(q,\kappa)}{8},
\label{2}
\end{equation}
where $q\rightarrow q_0=2N+1, N= 0,1,2,\cdots $ in the
case of $\kappa\rightarrow\infty$, i.e.  very high potential barriers
(i.e. harmonic oscillator approximation around a minimum
of the potential).  For barriers of finite height the parameter
$q$ is only approximately an odd integer $q_0$ in view
of tunneling effects.  The difference
$q-q_0$ is obtained by
imposing boundary conditions at
extrema of the potential (see below).
The second step is to insert (\ref{2}) into the equation, i.e. (\ref{1}),
and to write
\begin{eqnarray}
y&=&A(u) exp\bigg\{-\int\kappa sn u du\bigg\}\nonumber\\
&=&A(u) [f(u)]^{\kappa/2k}
\label{3}
\end{eqnarray}
where
$$
f(u) = \bigg(\frac{dn u + k cn u}{dn u - k cn u}\bigg).
$$
For large $\kappa$ the equation for $A(u)$ can be solved iteratively resulting
in an asymptotic expansion for $A(u)$ and concurrently one
for the remainder in eq. (\ref{2}), i.e. $\triangle $. A second
solution is written
\begin{equation}
y={\bar A}(u) exp\bigg\{+\int\kappa sn u du\bigg\}.
\label{4}
\end{equation}
The very useful property of these solutions is that for the
same value of $\triangle$ (which remains unchanged under
the combined replacements $q\rightarrow -q,
\kappa\rightarrow -\kappa$)
\begin{eqnarray}
{\bar A}(u) = A(u+2{\cal K}),\nonumber\\
{\bar A}(u,q,\kappa) = A(u, -q, -\kappa).
\label{5}
\end{eqnarray}
The domain of validity of these solutions
is that away from an extremum of the potential,
more precisely for
 $$
\bigg|\frac{dn u \mp cn u}{dn u \pm cn u}\bigg|>>\frac{1}{\kappa}.
$$
Thus one can construct solutions $Ec(u), Es(u)$, which
are respectively even in $u$ (or $sn u$) or odd, i.e.
\begin{equation}
\begin{array}{r } Ec(u) \\ Es(u)\end{array}
\propto A(u)[f(u)]^{\kappa/2k} \pm {\bar A}(u)[f(u)]^{-\kappa/2k}
\label{6}
\end{equation}
Since these expansions are not valid at the extrema of the
potential (where the boundary conditions are to be
imposed), one has to derive new sets of solutions there
and match these to the former
(i.e. determine their proportionality factors) in domains
of overlap (their extreme regions of validity).

Thus in the third  step two more pairs of solutions
$B, {\bar B}$ and $C, {\bar C}$ replacing $A, {\bar A}$ are
derived, one pair in terms of Hermite functions of a real variable, the other
in terms of  those of an imaginary variable, by transforming the
equations for $A,{\bar A}$ into equations in terms of
\begin{equation}
z(u)=\frac{\sqrt{8\kappa}}{k^{\prime}}\bigg(
\frac{dn u \mp cn u}{dn u \pm cn u}\bigg)^{1/2}, \;\; k^{\prime}=
\sqrt{1-k^2}.
\label{7}
\end{equation}
Solving the resulting equations iteratively as before, one obtains
again the same expansion for $\Lambda$, but solutions
$B,C$ replacing $A$, which are valid for
$$
\bigg|\frac{dn u \pm cn u}{dn u\mp cn u}\bigg|<<1.
$$
In their regions of overlap one can determine the proportionality
factors $\alpha, {\bar \alpha}$ of
\begin{equation}
B=\alpha A, \;\;\; {\bar C} = {\bar \alpha}{\bar A}
\label{8 }
\end{equation}
(again in the form of asymptotic expansions). Then
\begin{equation}
\begin{array}{r } Ec(u) \\ Es(u)\end{array}
\propto \frac{B[z(u)]}{\alpha}[f(u)]^{\kappa/2k} \pm
 \frac{{\bar C}[z(u)]}{{\bar \alpha}}[f(u)]^{-\kappa/2k}
\label{9}
\end{equation}

In the fourth  and final step one now applies the 
appropriate  boundary conditions
(cf. ref.\cite{8}) on these solutions (\ref{9}), i.e. one sets
\begin{eqnarray}
Ec(u=2{\cal K}) &=& Ec(u=0) = 0,\nonumber\\
Es(u=2{\cal K}) &=& Es(u=0) = 0,\nonumber\\
\bigg(\frac{\partial Ec}{\partial u}\bigg)_{u=2{\cal K}}
&=&\bigg(\frac{\partial Ec}{\partial u}\bigg)_{u=0} =0,\nonumber\\
\bigg(\frac{\partial Es}{\partial u}\bigg)_{u=2{\cal K}}
&=&\bigg(\frac{\partial Es}{\partial u}\bigg)_{u=0} =0.
\label{10}
\end{eqnarray}
These conditions define respectively functions of period $4{\cal K},
2{\cal K}, 2{\cal K}$ and $ 4{\cal K}$. Evaluating these one obtains
(from factors of factorials in $q$ and $-q$)  expressions
$\cot \{\pi(q-1)/4\}=\cdots, \tan\{\pi(q-1)/4\}=\cdots $,
from which the difference $q-q_0$ is obtained by
expansion around zeros. Finally
expanding
$$
\Lambda(q)\simeq \Lambda(q_0) + (q-q_0)\bigg(\frac{\partial\Lambda}{\partial q}
\bigg)_{q_0},
$$
one obtains the eigenvalues from which the level
splitting can be deduced.

In view of its periodicity the periodic potential avoids the
necessity of matching across turning points in the above calculation.
In the case of the double well potential this is
different, as explained in detail in ref. \cite{10}, and
one has to impose boundary conditions not only at the minimum but also
at the central maximum, again, of course, on even and
odd solutions constructed parallel to those above.

The perturbatively derived wave functions of ref.\cite{7}
(for large values of $\kappa^2$),
when matched in domains of overlap and so extended over the entire
domain of the variable $u$, and subjected to periodic
boundary conditions as described above  define two pairs of eigenfunctions,
in each case with one even and one odd, of periods $2{\cal K}$
and $4{\cal K}$ respectively. These four conditions together
imply for large values of $\kappa^2$ the following asymptotic
expansion of the eigenvalues $\Lambda$ as shown in ref. \cite{7}:
\begin{equation}
\Lambda_{\pm}(q_0)=\Lambda(q_0)\pm
\frac{2\kappa\bigg(\frac{2}{\pi}\bigg)^{\frac{1}{2}}}{[\frac{1}{2}(q_0-1)]!}
\bigg(\frac{1+k}{1-k}\bigg)^{-\frac{\kappa}{k}}\bigg(\frac{8\kappa}{1-k^2}
\bigg)^{\frac{1}{2}q_0}\bigg[1+O\bigg(\frac{1}{\kappa}\bigg)\bigg]
\label{11}
\end{equation}
Here $q_0=2N+1, N=0,1,2,\cdots$ and $\Lambda (q_0)$ is
the purely perturbative contribution which represents
effectively the eigenvalues of degenerate oscillators 
of the periodic potential in the case of very high barriers. 
It is the boundary conditions imposed on the perturbatively derived
solutions which yield the nonperturbative effects
equivalent to those of the instanton.  Thus the factor
$$
\bigg(\frac{1+k}{1-k}\bigg)^{-\kappa/k}
$$
is, in fact $exp(-S_0)$, where $S_0$ is the Euclidean action
of the instanton.

\vspace{0.2cm}

We first verify the result (\ref{11}) by reduction to the sine--Gordon
case. With $\kappa=\pm 2h$ finite
 while $n\rightarrow\infty$ and $k\rightarrow 0$
the Jacobian elliptic function $sn u$ reduces to $\sin u$ 
and eq.(\ref{1}) becomes by replacing $u$ by $x\pm \pi/2$
the Mathieu equation
\begin{equation}
y^{\prime\prime}+\bigg\{\lambda-2h^2\cos^2 2x\bigg\}y=0,
\;\; \lambda\equiv \Lambda-2h^2.
\label{12}
\end{equation}
Under the conditions stated the eigenvalues become
\begin{equation}
\lambda_{\pm}(q_0) 
=\lambda(q_0)
\pm\frac{4h(\frac{2}{\pi})^{\frac{1}{2}}(16h)^{\frac{q_0}{2}}}
{[\frac{1}{2}(q_0-1)]!}e^{-4h}\bigg[1+O\bigg(\frac{1}{h}\bigg)\bigg]
\label{13}
\end{equation}
in agreement with established results in this case \cite{9,10,11}. 

\vspace{0.2cm}

Next we consider $k$ approaching 1 in the case of the 
Lam${\acute e}$ eigenvalues (\ref{11})(terms up to  and including
those of $O(1/\kappa^2)$ in the level splitting and up
to and including those of $O(1/\kappa^4)$ in the perturbative
part have been given in ref. \cite{7} for any $q_0$).
One readily obtains
\begin{equation}
\Lambda_{\pm}(q_0)
=\Lambda (q_0)\pm\frac{(8\kappa)^{\frac{q_0}{2}+1}
(1-k)^{\kappa-\frac{1}{2}q_0}}
{[\frac{1}{2}(q_0-1)]!(2\pi)^{1/2}
2^{\kappa+1+\frac{q_0}{2}}}
\bigg[1+(1-k)\{\kappa(\frac{1}{2}-\ln 2)+\frac{q_0}{4}\}
+O\bigg(\frac{1}{\kappa}\bigg)\bigg]
\label{14}
\end{equation}
For the two lowest levels $q_0=1$ and one obtains
\begin{equation}
\Lambda_{\pm}(1)=\Lambda (1)
 \pm \frac{(4\kappa)^{3/2}(1-k)^{\kappa-\frac{1}{2}}}
{(2\pi)^{1/2}2^{\kappa}}\bigg[1+(1-k)\{\kappa(\frac{1}{2}-\ln 2)+\frac{1}{4}\}
+O\bigg(\frac{1}{\kappa}\bigg)\bigg]
\label{15}
\end{equation}
Thus the separation of the two lowest levels
is
\begin{equation}
\triangle\Lambda(1)\simeq\frac
{2(4\kappa)^{3/2}(1-k)^{\kappa-\frac{1}{2}}}
{{(2\pi)}^{1/2}2^{\kappa}}\bigg[1+(1-k)\{\kappa(\frac{1}{2}-\ln 2)+\frac{1}{4}\}
+O\bigg(\frac{1}{\kappa}\bigg)\bigg]
\label{16}
\end{equation}
This result agrees with formula (13) of ref.\cite{1} in the
limit of (in our notation) large $\kappa$ and $k\rightarrow 1$
(in particular this agrees also with our power
$\kappa-1/2$ of $(1-k)$ where in \cite{1}
the $-1/2$ has been ignored).
The higher order contributions are,
presumably, somewhat model or approximation dependent. We thus have
agreement with the instanton
result of ref.\cite{1} (there $\nu$ is our $k^2$, so that
for $k\rightarrow 1$ one has $(1-\nu)\rightarrow 2(1-k)$). 

Hence in the case of the periodic cosine potential,
as well as in the case of the Lam${\acute e}$ potential,
the method of matched asymptotic solutions  of refs.\cite{7,9} --
yields the same result as the instanton
method in the 1--loop
approximation, as one would expect.  In ref.\cite{1}
reference is made to the well known
case of the double well potential. 
For this case also it has been shown in ref. \cite{10}  that the method
of matched perturbation expansions  -- which one might argue
amounts to an improved form of the standard WKB method
(since it gives the correct eigenvalues) -- yields the
same result as the instanton method, again for any
arbitrary level, whether ground state or excited \cite{10}.

\section{Conclusions}
In the above we have demonstrated that
 the method of matched
asymptotic expansions of refs. \cite{7,9,10}
which, incidentally, was also developed and used to
determine the large order behaviour of the
perturbation expansion (cf. ref.\cite{13}),
 leads to the same result as the instanton
calculation in the 1--loop
approximation and thus is superior to a WKB
calculation.  The essential difference is
that WKB solutions alone,
or rather our WKB--like solutions involving the
function $A$, (which are not valid
at an extremum of the potential) do not suffice;
one has to match these to
solutions valid around the extrema.  
We may conclude that to obtain from the
Schr\"odinger equation and perturbation theory
 results agreeing with those of
the path integral method with expansion around the
instanton, one has to use the full 
method of matched asymptotic expansions
as developed in refs.\cite{9} and \cite{7,10}
(which has also been applied to other cases
such as spheroidal wave equations\cite{14}).
In purely quantum mechanical cases, such as those considered
above, the method of matched asymptotic 
expansions seems to be simpler and yields the
splitting of excited oscillator states with the same ease as 
that of the ground state, whereas in the case of the
path integral method, one has to use periodic instantons, as
 in e.g. refs. \cite{12,15}.
This may be worth noting in connection with
models of spin tunneling which attracted considerable interest
recently, since these can -- in certain cases and with
certain approximations (and coherent states)
-- be related to periodic differential equations.
Thus in ref.\cite{5} the case of the Hamiltonian
${\hat H} = K_1{\hat S}^2_z - K_2{\hat S}^2_x$
describing a ferromagnetic particle with large spin
has been considered and related to Mathieu and Lam${\acute e}$
equations. 
In particular the specific Lam${\acute e}$ instanton
of ref. \cite{1} has been used in ref.\cite{16}.
Finally we add that the introduction of the
parameter $q$ in conjunction with the construction of
solutions with the properties of eq. (\ref{5})
has been shown to be extremely useful in other
but related contexts as may be seen from the
calculation of the scattering matrix in ref.\cite{17}.

\vspace{1cm}

\noindent
{\bf Acknowledgement:} 
J.--z. Z. acknowledges support of the Deutsche Forschungsgemeinschaft
and Y. Z. that of an A. v. Humboldt--Foundation Fellowship.

\end{document}